\documentstyle[prl,aps,multicol,epsf]{revtex}
\renewcommand{\narrowtext}{\begin{multicols}{2}
\global\columnwidth20.5pc} 
\renewcommand{\widetext}{\end{multicols}
\global\columnwidth42.5pc} \multicolsep = 8pt plus 4pt minus 3pt

\begin{document}
\draft

\title{Spin degree of freedom in two dimensional exciton condensates}

\author{J. Fern\'andez-Rossier and C. Tejedor}
\address{Departamento de F\'{\i}sica Te\'orica de la Materia Condensada.
Universidad Aut\'onoma de Madrid. Cantoblanco, 28049 Madrid. Spain}  

\date{\today} 
\maketitle

\begin{abstract}

We present a theoretical analysis of a spin-dependent multicomponent 
condensate in two dimensions. 
The case of a condensate of resonantly photoexcited excitons having 
two different spin orientations is studied in detail. 
The energy and the chemical potentials 
of this system depend strongly on the spin polarization .
When electrons and holes are located in two different planes,
the condensate can be either totally spin polarized or spin unpolarized, 
a property that is measurable. The phase diagram in terms 
of the total density and electron-hole separation is discussed.

\end{abstract}


\narrowtext

In this letter we address the problem of a multicomponent quantum liquid: 
a condensate of excitons which hold an internal degree of freedom, 
namely the third component of the angular momentum $M$.\cite{spin}  
The multicomponent character is two-fold: on one side, the condensate is formed 
by pairs of distinct particles, an electron and a hole; on the other side, 
more than one kind of pair exists when $M$ is taken into account. From the 
later point of view, 
there is an evident analogy with some of the superfluid $^3$He phases in which 
a condensate with two components appears due to triplet pairing.\cite{leggett}
We consider excitons generated in a quasi two dimensional (2D) semiconductor 
heterostructures where carriers lifetimes can be very large. Moreover, 
degeneracy at the top of the valence band is broken usually leaving the highest 
hole with a third component of the angular momentum equal to $\pm 3/2$. 
Then, four types of excitons with 
different values of $M$ exist, implying that different components of the 
condensate with different populations $N_M$ for each spin are possible. 
The condensate properties depend on both the total density 
{\em and} relative populations. Hence, the problem offers more possibilities  
than the spinless exciton case considered in the 
literature;\cite{keldysh,jerome,comte,schmitt84,comte86,zhu,naveh} 
e.g. Josephson-type effects between the different components of the 
condensate\cite{mahan} are now possible.  

We have two main reasons for paying attention to $M$: 

i) The first is related to experiments. A promising way to obtain a 
multicomponent exciton condensate is to excite the semiconductor with   
circularly polarized light from a pulsed laser. 
Due to the angular momentum selection rule, two different kinds of excitons 
can be created, with either $M=+1$ or $-1$. The light polarization 
produces a spin polarization $P=(N_+-N_-)
/(N_++N_-)$\cite{note1}, where $N_\pm$ are the populations of excitons 
with $M=\pm 1$ respectively.\cite{damen,PRB2} 
It is interesting to study the properties of the photoexcited carriers 
with unbalanced populations $N_M$ by  
time resolved photoluminescence. There are time evolutions both  
on the total density due to recombination and on $P$ due to 
spin relaxation processes.  
Frequency and intensity of the emitted photons with the two different 
circular polarizations allow the experimental study of the condensates. 
It is possible to observe how the chemical potentials of each  
kind of excitons in the condensate are different; the splitting of 
the exciton levels turns out to be strongly dependent on $P$.\cite{PRB2} 

ii) The second reason is related to the physical origin of the existence of 
different phases and is precisely the aim of this letter. Each phase 
is characterized by different populations of each component of the condensate,
e. g. in the case of a condensate with two components with $M=+1$ and 
$M=-1$ respectively, it can be either a 
{\em ferromagnetic} phase with $N_+=N$ and $N_-=0$ or a 
{\em paramagnetic} phase with $N_+=N_-=N/2$.\cite{note1} 
The reason for the existence of different phases is the exclusion principle 
which plays a crucial role in the interaction between 
excitons in semiconductors. Effective masses are very light  so that 
electron and hole wave functions are very extended implying  
large overlaps which produce highly {\em spin dependent} effective 
interactions between excitons. We will show that the main  
consequence on that is that the ground state can be either  
spin polarized or spin unpolarized. This  depends on both the carrier density and  
the relative intensities of attractive and repulsive 
interactions that can be controlled by spatially separating electron 
and hole gasses with respect to each other. 

Our target is to study the ground state of $N$ electrons and $N$ holes 
lying in two different planes separated by a distance $d$. The Hamiltonian is  
given by
\begin{eqnarray}
\label{h1}
H&=& \! \sum_{\bf k \sigma } \! \left( 
\varepsilon_{\bf k }^e a_{\bf k \sigma }^\dagger a_{\bf k \sigma} + 
\! \sum _{{\bf k',q}, \sigma'} \frac{V_{\bf q}^{ff}} {2L^2} 
a_{\bf k \sigma}^\dagger a_{\bf k' \sigma'}^\dagger 
a_{{\bf k'}-{\bf q} \sigma'}a_{{\bf k}+{\bf q}\sigma} \! \right) \nonumber \\       
&+& \! \sum_{\bf k \varsigma } \! \left( 
\varepsilon_{\bf k  }^h b_{\bf k \varsigma}^\dagger b_{\bf k \varsigma} + 
\! \sum _{{\bf k',q}, \varsigma'} \frac{V_{\bf q}^{ff}} {2L^2} 
b_{\bf k \varsigma}^\dagger b_{\bf k' \varsigma'}^\dagger b_{{\bf k'}-{\bf q} 
\varsigma'} b_{{\bf k}+{\bf q} \varsigma}  \! \right) \nonumber \\  
&+&\sum_{\bf k,k',q,\sigma,\varsigma}
\frac{V_{\bf q}^{eh}(d)}{L^2} 
a_{\bf k \sigma}^\dagger b_{\bf k' \varsigma}^\dagger b_{{\bf k'}-
{\bf q}\varsigma} a_{{\bf k}+{\bf q}\sigma}  
\end{eqnarray}
where ${\bf k}$, ${\bf k}'$ and ${\bf q}$ are 2D wavevectors, 
$\sigma $ stands for the electron spin $\pm 1/2$ $(\uparrow, \downarrow )$   
and $\varsigma $ for the hole spin $\pm 3/2$ $(\Uparrow, \Downarrow )$ 
along the direction perpendicular to the 2D plane.  
Creation and annihilation operators are labelled as $a$'s for electrons and 
$b$'s for holes with single particle energies $\varepsilon _{\bf k}^e$
and $\varepsilon _{\bf k}^h$ respectively. $L^2$ is the area of the 
sample and the interactions $V_{\bf q}^{eh}(d)$ and $V_{\bf q}^{ff}$
($f \equiv e,h$) will be specified below. 

It is convenient to point out some assumptions behind this Hamiltonian. 
First of all, we suppose that the electron-hole ({\em e-h}) spatial degrees 
of freedom are in thermodynamic equilibrium. This hypothesis is supported 
by the fact that roughly 1 ps after the photoexciting laser pulse has finished,
the {\em e-h} liquid reaches a thermal momentum  
distribution.\cite{shah} That distribution corresponds to zero 
temperature when experiments are performed in the resonant excitation regime. 
Since recombination and spin relaxations are  
processes much slower than that thermalization,\cite{damen,PRB2} 
we take both the carrier density and the spin polarization as external 
parameters in the theory, in the spirit of a Born-Oppenheimer framework.

In order to look for exciton condensates we use a mean field 
Keldysh-Kopaev formalism, i.e. a generalization of the BCS 
treatment.\cite{keldysh,comte,schmitt84}
We define two propagators $G$ for each fermion,  
and four anomalous propagators $F$ for the four pairings
$(\uparrow,\Uparrow)$, $(\downarrow,\Downarrow)$, $(\uparrow,\Downarrow)$,
$(\downarrow,\Uparrow)$
\begin{eqnarray}
G_{\sigma}^e(k,t) & = & -i\langle T 
a_{\bf k,\sigma}(t)a_{\bf k,\sigma}^{\dagger}(0)\rangle \\ \nonumber 
G_{\varsigma}^h(k,t) & = & -i\langle T 
b_{\bf k,\varsigma}(t)b_{\bf k,\varsigma}^{\dagger}(0)\rangle \\ \nonumber 
F_{\sigma\varsigma}^{eh}(k,t) & = & -i\langle T 
a_{\bf k,\sigma}(t)b_{\bf k,\varsigma}(0)\rangle .
\end{eqnarray}
Each one of the anomalous propagators is associated with a different type  
of {\em e-h} pair $(\sigma ,\varsigma )$ and with a gap (order parameter
of the condensate) 
%\begin{eqnarray}
$\Delta_{\sigma \varsigma}(k)=\sum_{\bf q} V_{|\bf k-\bf q|}^{eh} 
\langle T a_{\bf q,\sigma}(0)b_{\bf q,\varsigma}(0)\rangle $
%\end{eqnarray}
proportional to the density of {\em e-h} pairs $(\sigma ,\varsigma)$. 
Following a standard mean field procedure one obtains the propagators:\cite{mahan} 
\begin{eqnarray}
G_{\sigma}^e (k,\omega)&=&\frac{ \Omega ^{h}_{\Downarrow} \Omega ^{h}_{\Uparrow}}
{\Omega ^{e}_{\sigma} \Omega ^{h}_{\Downarrow}\Omega ^{h}_{\Uparrow} 
+ \Omega ^{h}_{\Downarrow} \Delta_{\sigma \Uparrow}^2(k) + 
\Omega ^{h}_{\Uparrow} \Delta_{\sigma \Downarrow}^2(k)} 
\label{propag1} \\
F_{\sigma,\varsigma}^{eh} (k,\omega)&=&\frac{-\Delta_{\sigma 
\varsigma}(k) \Omega ^{h}_{-\varsigma}}
{\Omega ^{e}_{\sigma} \Omega ^{h}_{\Downarrow}\Omega ^{h}_{\Uparrow} 
+ \Omega ^{h}_{\Downarrow} \Delta_{\sigma \Uparrow}^2(k) + 
\Omega ^{h}_{\Uparrow} \Delta_{\sigma,\Downarrow}^2(k)}. 
\label{propag2}
\end{eqnarray}
For simplicity we have not written the hole propagator, $G_{\varsigma}^h 
(k,\omega)$, which has a similar expression to $G_{\sigma}^e (k,\omega)$ 
simply exchanging $e$ with $h$. $\Omega ^{h}_{\varsigma}(k,\omega)=-i \omega 
-(\varepsilon_{\bf k }^h -\mu_{\varsigma}^{h}-\Sigma_{\varsigma}^{h}(k))$, 
$\Omega ^{e}_{\sigma}(k,\omega)=-i 
\omega +\varepsilon_{\bf k }^e -\mu_{\sigma}^{e} -\Sigma_{\sigma}^{e}(k)$ with  
$\mu _\sigma ^e$ and $\mu _\varsigma ^h$ being electron and hole chemical 
potentials  and $\Sigma$ is the usual Hartree-Fock fermion self-energy. 

In the expressions for the propagators one observes the difference between the 
condensate we are studying and those occurring in other problems like 
superconductors, $^3$He superfluid or the spinless exciton gas. Since the pairing 
occurs among two different types of fermions, each one with different 
spin, the terms $\Omega ^e_\sigma \Omega ^h_{\varsigma }\Omega ^h_{-\varsigma }$ 
appearing in the denominator produce three poles for each propagator. 
This causes a complicate mixing of excitations 
unlike other types of condensates.\cite{future1} 

In this paper we will focus on the case of resonant excitation 
of $(\downarrow,\Uparrow)$ pairs. Under resonance conditions, 
holes do not relax their spin indepently\cite{damen,maialle}, a process 
which whould produce $(\downarrow,\Downarrow)$ pairs with $M=-2$. 
Instead, the main effect is a relaxation of the angular momentum of the 
whole exciton, by means of an intra-exciton exchange spin-flip,  
from $M=+1$ to $M=-1$.\cite{damen,maialle} Therefore, we neglect the   
densities of excitons with both $M= \pm 2$ 
and only two kinds of excitons are considered: 
those with $M=-1\equiv (\uparrow,\Downarrow)$ 
and those with $M=+1\equiv (\downarrow,\Uparrow)$
having densities $n_{-}=N_-/L^2$ and $n_{+}=N_+/L^2$, respectively.
Such a system will be characterized by two quantities, 
the total density $n=N/L^2$ and the polarization $P$. 
In this case, either the second or the third term in the denominators 
of Eqs.(\ref{propag1}),(\ref{propag2}) disappears, one of the $\Omega ^h $'s 
can be cancellated out and each propagator only has two poles as in the BCS or 
superfluid $^3$He theory. Then one could follow standard procedures\cite{mahan} 
to obtain gap equations for $\Delta _{\downarrow, \Uparrow}(k)$ and  
$\Delta _{\uparrow,\Downarrow}(k)$ 
(hereafter denoted as $\Delta _+(k)$ and $\Delta _-(k)$ respectively)  
which are coupled to each other solely through the Hartree contribution which is 
spin independent.\cite{future1} For an homogeneous system, the Hartree term 
can be embedded in the chemical potential as an energy origin so that 
the gap equations become decoupled to all the effects. Then, solving 
these gap equations is equivalent to looking for an extremal from 
\begin{equation}
\langle \Psi | H- \mu_{+} N_{+}-\mu_{-} N_{-}| \Psi \rangle, 
\label{functional}
\end{equation}
using a wave function 
\begin{equation}
|\Psi \rangle \! = \! \prod_{\bf k,\bf k'} \! \left(u^{+}_{k}+v^{+}_{k} 
a_{\bf k \downarrow}^\dagger b_{\bf k \Uparrow}^\dagger \right) 
\! \left(u^{-}_{k'}+v^{-}_{k'} a_{\bf k' \uparrow}^\dagger 
b_{\bf k' \Downarrow}^\dagger \right) \! \mid 0 \rangle . 
\label{Wavefunc.}
\end{equation}
$\mu _{\pm}$ are the chemical potentials of $\pm 1$ excitons and 
$\mid 0 \rangle $ is the ground state of the system without any 
photoexcited {\em e-h} pairs. 
Each parenthesis in Eq. (\ref{Wavefunc.}) corresponds to one of the 
two components of the condensate. 
As in the spinless case, this wave function interpolates smoothly between 
two limits, $n \rightarrow 0$ and $n \rightarrow \infty $, where 
it is exact.\cite{comte} The normalization condition 
$\langle \Psi\mid\Psi\rangle =1$ of the condensate wave function 
imposes the constraints $(u_{k}^{\pm})^{2}+ (v_{k}^{\pm})^{2} =1$. 
The trial wave function (\ref{Wavefunc.}) is valid under resonance 
conditions, as discussed above, and it leads to a gap equation 
$\Delta_{\pm}(k)=\sum_{\bf q} V_{| \bf k- \bf q|}^{eh} u^ \pm _q v^\pm _q$.
The practical procedure is to fix the two chemical potentials 
$\mu _\pm $ and to minimize the functional (\ref{functional}) in terms of 
the parameters $u_{k}^{\pm}$, $v_{k}^{\pm}$ in a discretized $k$-space. 
Since the Coulomb interaction is singular in 2D, we follow the ususal 
procedure \cite{schmitt84} of adding and subtracting a term that 
makes the singularity analytically integrable. 
From $u_{k}^{\pm}$, $v_{k}^{\pm}$ we get $\Delta_{\pm}(k)$ and 
the density, $n_{\pm}=\sum _k (v^\pm _k)^2$, for each type of {\em e-h} pair.  
So we have all the ingredients to obtain the Green functions, 
the energy per pair, etc.. Changing $\mu_{\pm}$ we can obtain the 
dependence of the energy per pair for each spin on both $n$ and $P$. 
The results so obtained for $\mu_\pm $ as a function of $n$ and $P$ are directly 
comparable with photoluminescence experiments which measure the 
energy liberated when an exciton recombines.\cite{damen,PRB2}  

Let us discuss the physics involved in our mean field theory. 
We will not include screening which does not depend on $P$. When 
treated to the lowest order in perturbation theory,\cite{joaquin} it does 
not affect the splitting between chemical potentials $\delta 
=\mu_{+}-\mu_-$ which is our main interest in this paper. 
In this approach there are three many-body contributions 
to the renormalization of the single exciton chemical potential: 
The first is a Hartree term which gives 
the unique coupling between the two components of the condensate. The other two  
contributions are related to the Pauli principle. 
One is the exchange correction (EC), i.e. the reduction 
of the repulsive electron-electron (hole-hole)  
interaction due to the fact that two identical fermions can not occupy the 
same state. The last contribution is the vertex correction (VC), which represents  
the reduction of the {\em e-h} attraction due to the occupation of  
final states in {\em e-h} scattering processes. 
EC and VC contributions to the chemical potential of   
a fermion with a given spin are independent of 
the amount of fermions with the opposite spin. 
This causes a non zero splitting, $\delta $, 
whenever there is a non zero spin polarization $P$,  
in agreement with experiments.\cite{damen,PRB2}  
All these physical properties are described by a wave function (\ref{Wavefunc.}) 
for the condensate which is separable in two components in analogy what 
happens in the A phase of superfluid $^3$He due to {\em equal spin pairing} 
(ESP) states.\cite{leggett} The exciton 
condensate that we present in this paper is more flexible than ESP 
states of $^3$He because $P$ and $\delta$ can be non zero in the exciton case, 
while they must be zero in the ESP case.   
 
The numerical results shown in this paper are obtained with 
$V_{\bf q}^{eh}(d)=-e^2 exp(-qd) /\epsilon q$, i.e.  
the Fourier transform of $V^{eh}(r,d)=-e^2/ \epsilon \sqrt{r^2 +d^2}$, and 
$ V{\bf _q}^{ff}=-V_{\bf q}^{eh}(d=0)$, 
$\epsilon$ being the dielectric constant of the material. 
The single particle energy bands are supposed to be isotropic and parabolic.
The spinless case has been studied 
by Zhu {\em et al.}\cite{zhu}, who considered $d\neq0$ because the  
{\em e-h} separation increases the exciton lifetime 
improving the condensation conditions and avoids formation of biexcitons. 
We have an additional reason to be interested in the case $d\neq0 $: 
$d$ is a parameter which controls the relative strengths of   
EC and VC which compete with each other. 
Hence, by tuning $d$ and $n$, one can obtain a quantum phase transition 
between a ferromagnetic phase (dominated by exchange) and a paramagnetic phase 
(dominated by the vertex corrections). 
This is a crucial result coming out from our study of the spin degree of 
freedom in the exciton condensate.

For all the results presented in this paper we scale the length to the  
three-dimensional Bohr radius $a=\hbar^2 \epsilon /m^{\ast }e^2$, where 
$m^{\ast }=m^{e}m^{h}/(m^e +m^h)$ is the exciton reduced mass. 
Energies are taken with origin at the bottom of the conduction band and  
are given in units of in the 2D Rydberg $|E_{0}^{2D}|=2e^2/ \epsilon a$. 

Figure 1 shows chemical potentials $\mu _\pm $ as a function of the  
density for $P=0.8$ and three values of the separation $d$. 
$\mu _\pm $ are always increasing with $n$ because exciton-exciton interaction 
turns out to be globally repulsive. That means that repulsive Hartree and VC terms 
in the energy dominate on the attractive EC. The essential point is the 
behavior of $\delta =\mu _+ - \mu _-$, which is a decreasing function 
of $d$ for moderate densities and it can be positive or negative.  
This is shown more clearly in the inset. The first conclusion is that 
$\delta$ is in the range of meV and is consequently an experimentally accessible  
quantity.\cite{damen,PRB2} 
Since the Hartree term is spin independent, it does not contributes to 
$\delta $ which only depends on EC and VC.   
As the {\em e-h} separation increases, the VC decreases while the EC 
increases as a consequence of a larger exciton size that is implied 
by the increase of the overlap between excitons. Therefore, for moderate 
densities there is, for a critical value 
$d_{cr}\simeq 0.12 a\simeq 10 \AA$, a change of sign in the measurable 
quantity $\delta $. Within our numerical precision, 
this critical separation turns out to be independent on $P$. 
For $d<d_{cr}$ VC dominates on 
EC giving $\delta > 0$ while for $d>d_{cr}$ EC dominates and $\delta <0$. 
It must be stressed that the value of $d_{cr}$ is 
quite low and the change of sign of the splitting effect could be 
observable, for example, applying moderately high external electric fields.  

Figure 2 shows the phase diagram of the sign of $\delta $ in terms of 
the physical magnitudes of interest in the problem: $na^2$ and $d/a$.
For high $d$ or $n$ the kinetic energy dominates the {\em e-h} 
attraction and the {\em e-h} pairs are not bound having a 
positive chemical potential. 
Therefore, experimental determination of the splitting must be performed 
in the more accessible region of low $d$ and $n$ where excitons are well defined. 
The important point is that the diagram also gives information about the  
polarization of the condensate. This is easily understood by realizing that 
the derivative of the total energy per pair $E$, in the thermodynamic limit, is: 
\begin{eqnarray}
\left( \frac{\partial E}{\partial P}\right) _N \equiv E(N_++1,N_--1)-E(N_+,N_-)
=\delta .
\label{deriv}
\end{eqnarray}
Since for fixed $d$ and $n$, $\delta $ does not change sign when $P$ varies, 
the derivative (\ref{deriv}) is a monotonous function of $P$. 
The region of the diagram with positive splitting 
corresponds to an increase of the total energy $E$ with $P$, 
i. e. the system has minimum energy when it is unpolarized, $P_{me}=0$. 
However, in the region with $\delta <0$, the derivative 
(\ref{deriv}) is also negative indicating that the system should have 
minimum energy when completely polarized, $P_{me}=1$.   
For $d<d_{cr}$ VC always dominates and the unpolarized state $\delta >0$ 
has the minimum energy. For increasing separation $d$, the excitons are less 
strongly bound and EC dominates allowing the appearance of a polarized 
phase with $\delta <0$. It is difficult 
to predict whether, for adequate values of $d$ and $n$, the spins 
of the condensate would spontaneously polarize or not because this 
process should occur in the presence of spin relaxation mechanisms that 
we have not considered, as mentioned above. 
Therefore, the best way of experimentally analyzing the phase diagram is 
to measure the sign of $\delta $ instead of looking for magnetic properties.  
$\delta $ would remain as given by Fig. 2 
even if other physical contributions were to be included in our analysis. 
Available experiments\cite{damen,PRB2} are for $d=0$ and, in the whole 
range of experimentally accessible densities, always 
$\delta $ results to be positive as it happens in our phase diagram.

In summary, we have studied a condensate of 2D excitons 
having different spin orientations. We have shown that the spin 
degree of freedom plays a very important and non-trivial role in 
exciton condensation. We have analyzed the case of resonant excitation 
finding that the ground state energy depends 
strongly on spin polarization. When electron and hole components are  
separated from each other, the relative importance of exchange and 
vertex correction interactions can be altered. This causes a 
change of sign in $\delta $. This 
splitting between the chemical potentials of different exciton components can 
be measured by time-resolved spectroscopy.\cite{damen,PRB2} 
The sign of $\delta $ is related to the minimum energy configuration 
of the condensate, i.e. spin polarized or spin unpolarized. 
All these features are shown and discussed in a phase diagram in $n$ and $d$.

We thank L. Vi\~na and L. Brey for useful discussions. Work  
supported in part by CICYT of Spain under contract No. MAT 94-0982-C02-01
and by the Comunidad Autonoma de Madrid under contract AE00330/95.

\begin{figure}
\caption{Majority $\mu _+$ (continuous lines) and minority $\mu _-$ 
(dashed lines) chemical potentials as 
a function of the density $n$ for $d=0.0$, $0.15$ and $0.30$. 
The inset shows the splitting $\delta =\mu_{+}-\mu_-$ as a function of 
the {\em e-h} separation $d$ for $na^2=0.15$ and $P=0.5$ and $0.8$.}
\label{fig1}
\end{figure}

\begin{figure}
\caption{Phase diagram for both sign of $\delta =\mu_+ -\mu _-$ and 
polarization $P_{me}$ of the condensate (see text). 
Shaded region corresponds to a condensate of unbounded {\em e-h} pairs.}
\label{fig2}
\end{figure}

\widetext

\begin{references}
\bibitem{spin} The third component of the angular momentum $(M)$ 
is usually denoted in the field as the {\em spin} . 
\bibitem{leggett}A.J. Leggett, Rev. Mod. Phys. {\bf 47}, 331 (1975). 
\bibitem{keldysh} L.V. Keldysh and Yu.V. Kopaev, Fiz.\ Tverd.\
Tela {\bf 6}, 2791 (1964) [Sov.\ Phys.\ Solid State {\bf 6}, 2219
(1965)].
\bibitem{jerome} D. Jerome {\em et al.}, Phys. Rev.
{\bf 158}, 462 (1967); B.I. Halperin and T.M. Rice, Solid State Phys. {\bf 21}, 
(Academic Press, New York, 1968) p. 115.; 
R. Zimmermann, Phys. Status Solidi B {\bf 76}, 191 (1976). 
\bibitem{comte}C. Comte and P. Nozi\`{e}res, J. Physique {\bf 43}, 1069 (1982); 
P. Nozi\`{e}res and C. Comte,{\em ibid}  {\bf 43}, 1083 (1982).
\bibitem{schmitt84}H. Haug {\em et al.}, Prog. Quant. Electr. {\bf 9}, 3
(1984).
\bibitem{comte86}C. Comte and G. Mahler, Phys. Rev. B {\bf 34}, 7164,(1986).
\bibitem{zhu}X. Zhu {\em et al}, Phys. Rev. Lett. {\bf 74}, 1633 (1995).
\bibitem{naveh}  X. Naveh {\em et al.}, Phys. Rev. Lett. 
{\bf 77}, 900 (1996); 
T. Portengen {\em et al.}, Phys. Rev. Lett. {\bf 76}, 3384 (1996);
H. Chu and Y.C. Chang Phys. Rev. B. {\bf 54}, 5020 (1996).
\bibitem{mahan} G. Mahan, {\em Many Particle Physics}, (Plenum, New York, 1982).
\bibitem{note1} Along this paper we take $N_+ \geq N_-$.
\bibitem{damen} T.C. Damen {\em et al.}, Phys. Rev. Lett. {\bf 67}, 3432 (1991).
\bibitem{PRB2} L. Vi\~na, {\em et al.}, Phys. Rev. B {\bf 54}, R8317 (1996). 
\bibitem{shah} J. Shah, {\em Hot Carriers in Semiconductor Nanostructures}, 
(Academic Press, San Diego, 1992). 
\bibitem{future1} J. Fern\'andez-Rossier and C. Tejedor, (unpublished).
\bibitem{maialle} M. Z. Maialle {\em et al.}, Phys. Rev. B {\bf 47}, 15776 (1993). 
\bibitem{joaquin} J. Fern\'andez-Rossier {\em et al.}, Phys. Rev. B
{\bf 54}, 11582 (1996).
\end{references}
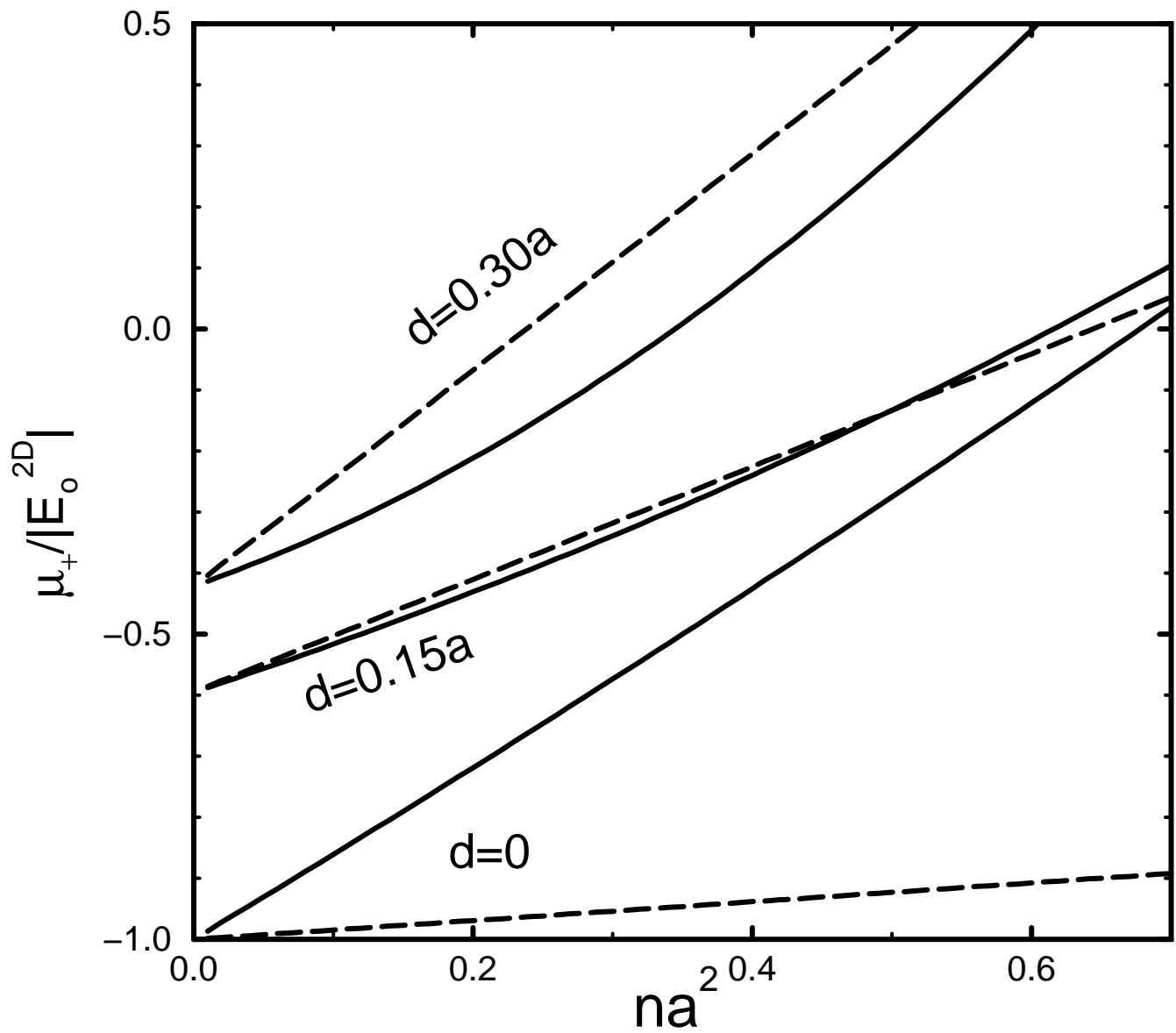
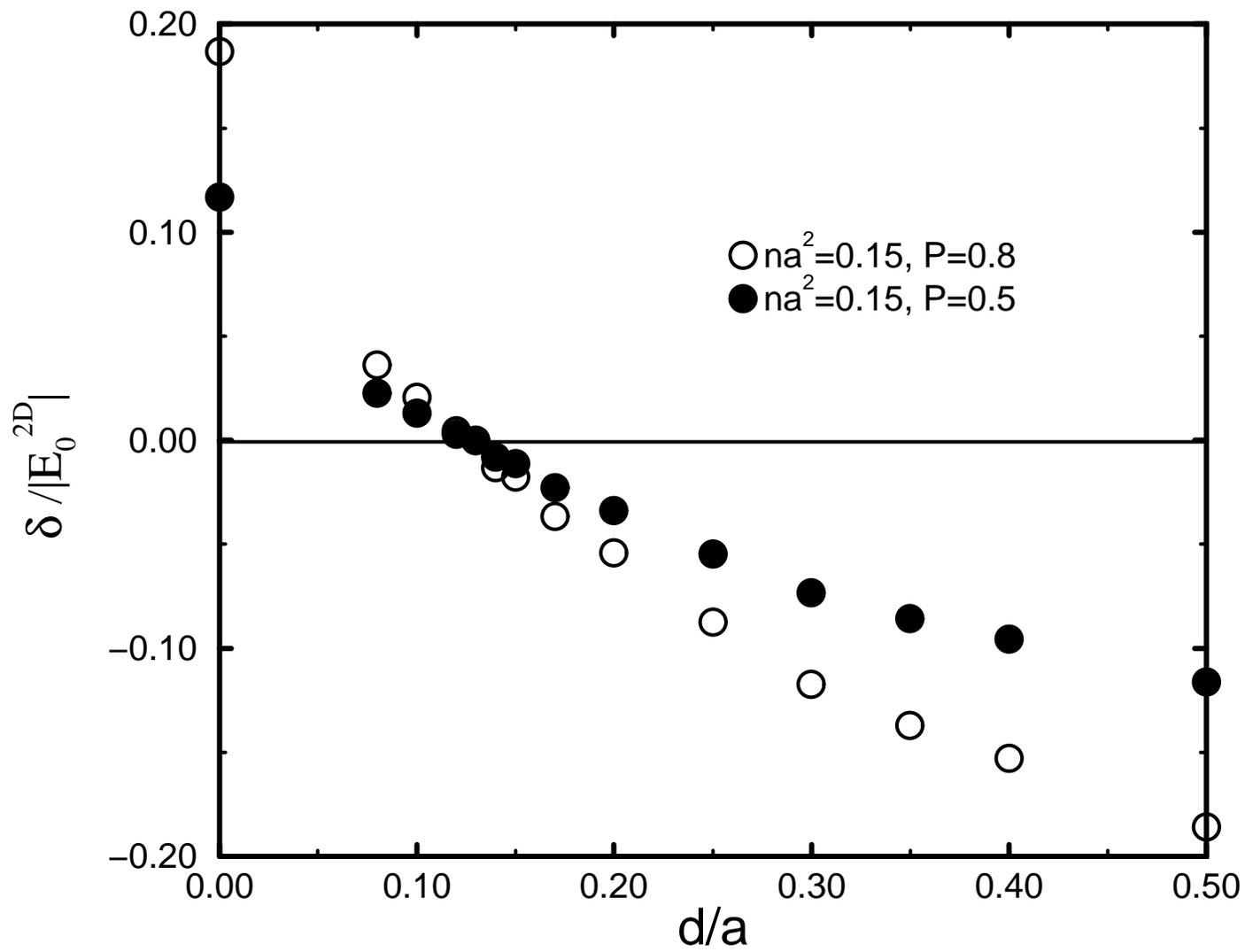
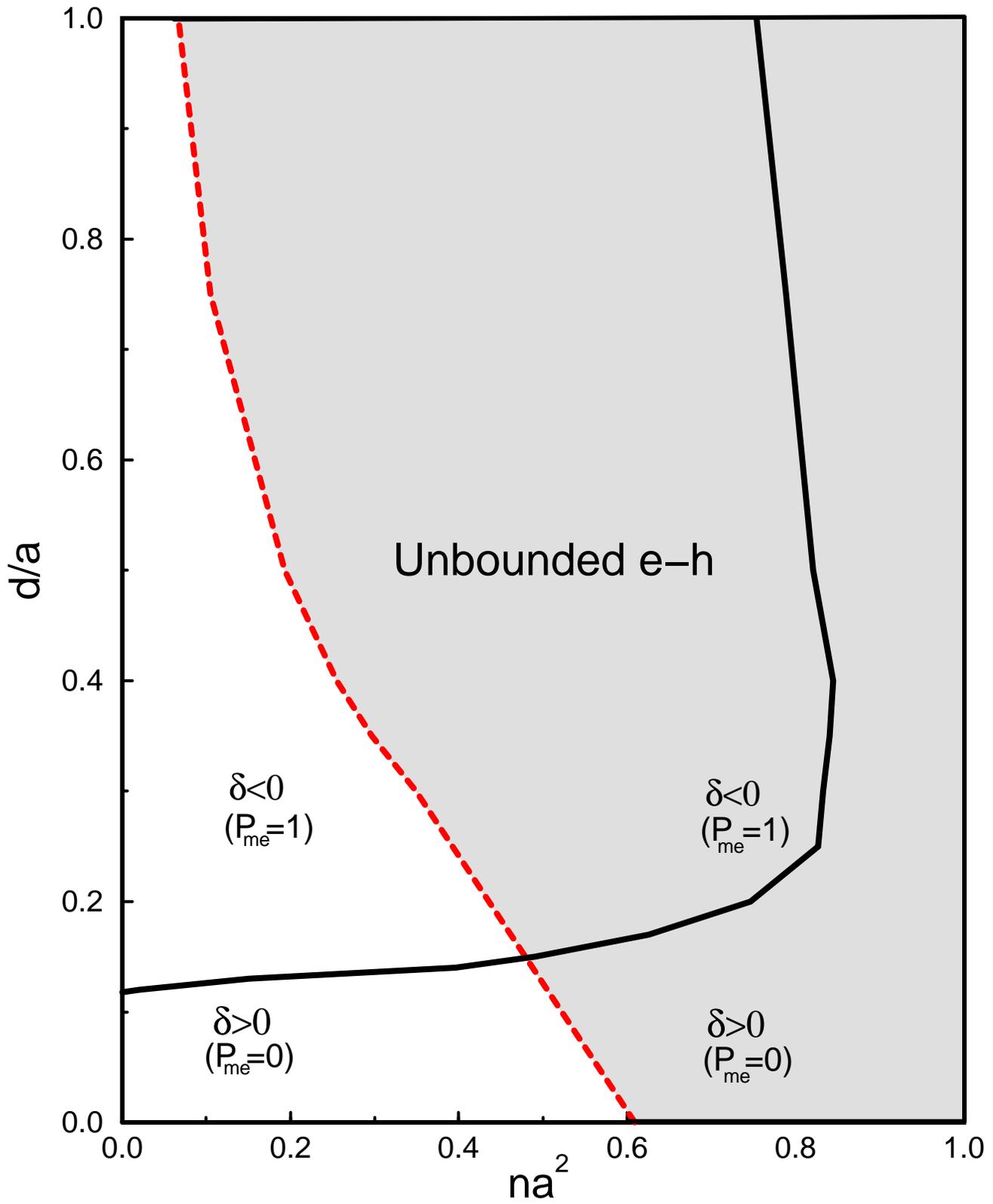
\end{document}